\documentclass[aps,prl,reprint,superscriptaddress,10pt,twocolumn,amsmath,amssymb,showpacs]{revtex4-1}
\usepackage{graphicx,wasysym,color,SIunits,chemformula}
\usepackage[bottom]{footmisc}

\usepackage{chemist}

\newcommand{\del}[1]{}

\usepackage{epsfig}
\usepackage{epstopdf}
\usepackage{float}
\usepackage{color}
\usepackage[latin1]{}
\usepackage[normalem]{ ulem }
\usepackage{soul}
\usepackage{newtxmath}

\begin{document}

\title{Active pulsatile gels: from chemical microreactor to polymeric actuator}

\author{Baptiste Blanc}
\affiliation{Department of Physics, Brandeis University, 415 South St., Waltham, MA 02454, United States}
\author{Zhenkun Zhang}
\affiliation{Key Laboratory of Functional Polymer Materials of Ministry of Education, Institute of Polymer Chemistry, College of Chemistry, Nankai University, 300071 Tianjin, China}
\author{Eric Liu}
\affiliation{Department of Chemical and Biological Engineering, Tufts University, Medford, Massachusetts 02155, United States}
\author{Ning Zhou}
\affiliation{Department of Chemistry, Brandeis University, 415 South St., Waltham, MA 02454, United States}
\author{Ippolyti Dellatolas}
\affiliation{Department of Mechanical Engineering, Massachusetts Institute of Technology, Cambridge, Massachusetts, 02139 USA}
\author{Ali Aghvami}
\affiliation{Department of Physics, Brandeis University, 415 South St., Waltham, MA 02454, United States}
\author{Hyunmin Yi}
\affiliation{Department of Chemical and Biological Engineering, Tufts University, Medford, Massachusetts 02155, United States}
\author{Seth Fraden}
\affiliation{Department of Physics, Brandeis University, 415 South St., Waltham, MA 02454, United States}

\begin{abstract}
We report on a synthesis protocol, experimental characterization and theoretical modeling
of active pulsatile Belousov-Zhabotinsky (BZ) hydrogels. Our two-step synthesis technique allows independent optimization of the geometry, the chemical, and the mechanical properties of BZ gels. We identify the role of the surrounding medium chemistry and gel radius for the occurrence of BZ gel oscillations, quantified by the Damk\"{o}hler number, ratio of chemical reaction to diffusion rates. Tuning the BZ gel size to maximize its chemomechanical oscillation amplitude, we find that its oscillatory strain amplitude is limited by the timescale of gel swelling relative to the chemical oscillation period. Our experimental findings are in good agreement with a Vanag-Epstein model of BZ chemistry and a Tanaka Fillmore theory of gel swelling dynamics.

 \pacs{.}
\end{abstract}

\maketitle
\section{Introduction}
The existence of living organisms is evidence that functional motile autonomous materials can be built robustly using macromolecules. However, engineering polymeric materials that possess these properties remains a longstanding goal, with sought-after applications in infrastructures inspections, programmable matters and smart medecine \cite{Li2021(2),Liu2021}. Current endeavors in this field can be classified in two main categories. The first one consists in constructing life-like materials by assembling individual chemomechanical building blocks borrowed from the living realm such as molecular motors, enzymes, or cells because these elements have the coveted capability of autonomously converting chemical energy into mechanical work \cite{Wu2017,Litschel2018,Testa2021,Wang2020}. The second one couples passive engineered polymeric materials such as hydrogels, colloids or elastomers to external power sources \cite{Wehner,Christianson,Shin}. While both of these approaches have shown promising results, none of them tackle the challenge of building from the molecule to the macroscopic scale a purely synthetic functional chemomechanical materials that harness chemical energy to produce motion. This scientific effort is in its infancy and could lead to the creation of life-like machines with augmented performance compared to what evolution has created \cite{Katsonis,Rothemund2021}. 
\par
Inspired by Yoshida's work \cite{Yoshida1996(2)}, we create synthetic autonomous chemomechanical polymeric materials made of acrylamide hydrogels doped with the catalyst of the Belousov Zhabotinsky(BZ) reaction \cite{Zhabotinsky1964} . In the BZ reaction, an autocatalytic activator drives the catalyst to its oxidized state and generates an inhibitor that returns the catalyst to its reduced state \cite{Noyes1972}. The cyclic change of the oxidation state of the catalyst induces the oscillatory swelling of the gel.
Controlling the strain of BZ gels is crucial for the design of shape-changing functional elements such as bimorphs that rely only on a bending deformation \cite{Na2015,Levin2020,Li2021}. Yet maximizing the BZ gel oscillatory strain amplitude is still an open challenge for material scientists. Over the past 25 years, multiple parameters such as the composition of the BZ gels \cite{Aizenberg2018,Kramb2015,Masuda2015,Wang2022}, the architecture of the catalyst \cite{Aizenberg2018,Zhang2013}, the charge of the monomer \cite{Maeda2007,Aizenberg2018}, the porous structure of the hydrogel \cite{Aizenberg2018,Takeoka2003,Suzuki2012,Hu2017}, its shape \cite{Yuan2013,Chen2011} and its chemical environment \cite{Buskohl2016} have been reported to influence the swelling of a BZ gel, which testifies to the complexity of this chemomechanical process. Moreover, each of these studies uses a different fabrication process. This variability in both observations and experimental protocols calls for the design of a modular platform of BZ gel fabrication and characterization. Such advances would allow for an understanding of the physicochemical phenomena at play in this specific BZ material, and also would help establish rules that are general to any hydrogel-based actuators coupled to a chemical oscillator \cite{Litschel2018,Cangiolosi2017}.
\par
In this work, we first develop a platform for the rational design of BZ chemomechanical gels, which separates the hydrogel fabrication from its functionalization with the BZ catalyst. We demonstrate the modularity of our synthesis protocol by creating spherical BZ gels of tunable chemical composition and size. With this isotropic material, we highlight the role of the size of the gel in its chemical behavior. Chemical oscillations are only observed in spherical BZ gels with a radius larger than a critical size. This result can be rationalized by estimating the Damk\"{o}hler number, which compares the production rate of the activator of the BZ reaction inside the gel to its loss rate through diffusion in the outer solution. We then show experimentally that the timescale of BZ gel swelling increases with the square of its radius. For the smallest BZ gel above the critical size, expected to have the largest oscillatory strain, we find that its strain reaches only $5\%$ of its theoretical maximum.  These results demonstrate that the factor limiting its oscillatory strain amplitude is the long timescale of gel swelling compared to the shorter period of the BZ oscillation. A strategy to overcome this impediment is suggested.\par

\section{Results and discussion}

\subsection{Two step synthesis of BZ gels}
\begin{figure*}
\centering
\includegraphics[width=0.95\textwidth]{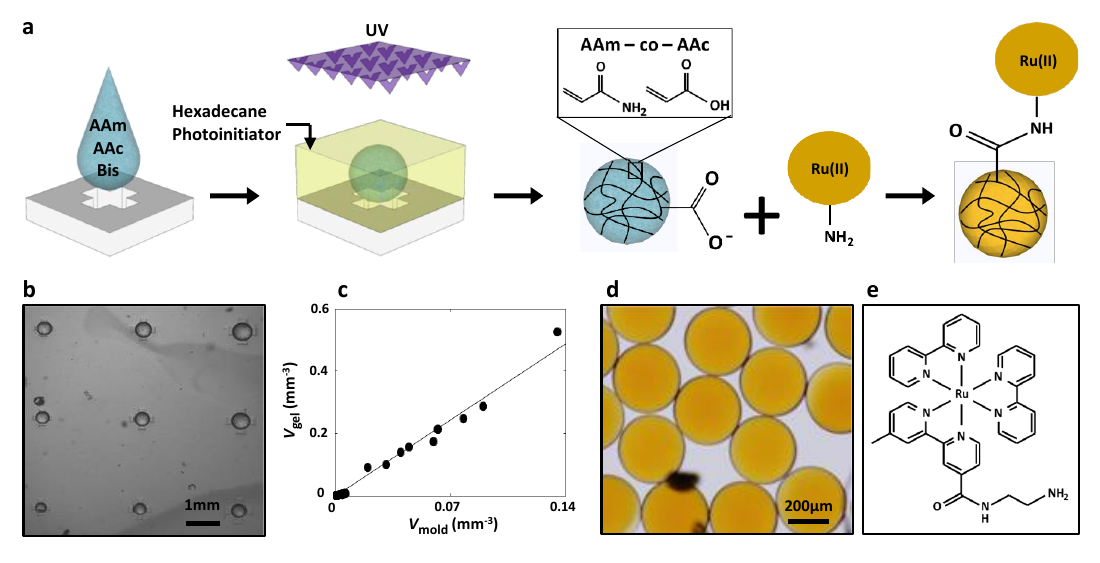}
\caption{(a) Schematic of the gel synthesis process. (b) Bright-field image of pre-gel drops on a PDMS mold before UV exposure. The drops are an aqueous solution of 15$\%$ acrylamide (AAm), 1.5$\%$ acrylic acid (AAc), 1$\%$ bisacrylamide (Bis). The outer solution is hexadecane. (c) Relation between the volume of the micropatterned wells of the PDMS mold and the volume of the crosslinked gel. The line represents the theoretical prediction assuming that the volume of the gel is proportional to the volume of the micropatterned wells of the PDMS. (d) Bright-field image of 200 $\mu$m diameter gels of composition 15$\%$ $\mathrm {AAm}$, 1.5$\%$ $\mathrm {AAc}$, 1$\%$ $\mathrm {Bis}$ coupled with the ruthenium catalyst. (e) Chemical structure of the ruthenium catalyst synthesized in this study:  Bis(2,$2^{'}$-bipyridine)-$4^{'}$-methyl-4-amido ethylamine-2,$2^{'}$-bipyridine-ruthenium bis(hexafluorophosphate), noted $\mathrm{Ru(bpy)_{2}(bpyEtAm)(PF_{6})_{2}}$.}
\label{figure1}
\end{figure*}

BZ hydrogels are fabricated through a two-step process (Figure \ref{figure1}-a)\cite{Masuda2015,Smith2012,Kramb2014}. First, we create a hydrogel by simultaneously polymerizing mixtures of inert and functional monomers in predetermined stoichiometric ratios and crosslinking the polymers together. Second, we attach a pendant BZ catalyst to the functional groups. This two-step process is designed to allow the production of BZ gels with independent control over their mechanical properties, their chemically oscillating behavior and their geometry. The stiffness of the gel can be systematically controlled by varying the cross-link density before introducing the catalyst, while the catalyst concentration is set by the concentration of functional monomers. Control over the shape and size of BZ gels can be accomplished by using available molding, 3D printing, patterning or self assembly techniques.\par
Here, we fabricate isotropic and homogeneous spherical BZ hydrogels. Spheres are chosen to limit the complexity of the chemomechanical process resulting from any shape anistropy of a BZ hydrogel. We combine an emulsion polymerization approach and a molding process to create hydrogel spheres from 100 $\mu$m to 600 $\mu$m in diameter (Figure \ref{figure1}-a-b-c)\cite{Liu2016}. Acrylamide, bisacrylamide and acrylic acid are used as the main monomer, the crosslinker and the monomer holding the carboxilic acid functional group respectively. We vary the crosslinker concentration in the prepolymer solution from $0.05\%$ (weight per volume (w/v)) to $1\%$ (w/v) to control the hydrogel stiffness. We could not obtain crosslinked gels for a concentration of crosslinker below $0.05\%$ (w/v). The acrylic acid concentration is tuned from $1\%$ (w/v) to $4\%$ (w/v) in order to change the concentration of carboxilic groups in the hydrogel. Further experimental details on the spherical hydrogel fabrication can be found in Experimental Section-2. We then covalently bind a home made ruthenium catalyst containing an amine group to the carboxilic groups of the spherical hydrogel through carbodiimide crosslinker chemistry (Figure \ref{figure1}-a-d-e). The synthesis protocol for the ruthenium catalyst, the protocol and the proofs for its covalent binding to the hydrogel are described in Experimental Section 3 and 4.\par
With this technique, we report homogeneously incorporating synthetized $\mathrm{Ru(bpy)_{2}(bpyEtAm)(PF_{6})_{2}}$ in spherical BZ gel with catalyst concentrations up to 12 mM, which is significantly higher that the solubility limit in water. Notable desirable features of this two-step process are the ability to: 1) synthetize BZ hydrogels of spherical shape which impart them isotropic properties, 2) create BZ hydrogels smaller than the size of the oxidizing front of the BZ reaction ($\sim$500 $\mu$m), allowing to work in a regime where the oxidation state of the BZ gel can be considered homogeneous, 3) homogeneously incorporate the ruthenium trisbipyridine catalyst in the BZ hydrogels, 4) create BZ hydrogels with a higher catalyst concentration than the solubility limit, 5) create BZ hydrogels with tunable stiffness. 

\subsection{Influence of hydrogel size on BZ oscillatory chemistry}
\par
\begin{figure}
\centering
\includegraphics[width=0.48\textwidth]{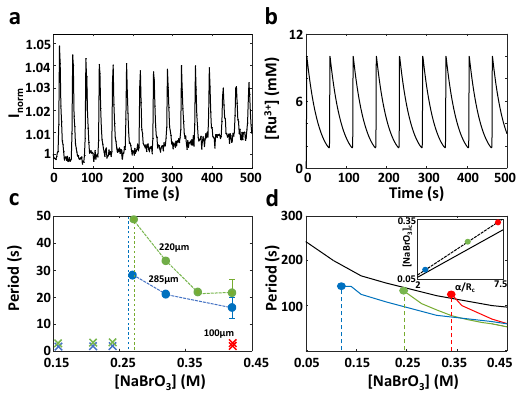}
\caption{(a) Intensity profile of a 220 $\mu m$ radius BZ gel bead as a function of time in a BZ solution with a concentration of sodium bromate of 0.42 M. (b) The catalyst concentration as a function of time obtained from a Vanag Epstein model with a diffusive flux coefficient $\it{\frac{\alpha}{R}}$=5 $s^{-1}$ and with a concentration of sodium bromate of 0.42 M. The model shows steady oscillation. (c) Experiment. Period of the BZ gel chemical oscillation as a function of the bromate concentration for different size of beads. Gels of 100 $\mu m$, 220 $\mu m$, 285 $\mu m$ radius are represented respectively in red, green and blue. Disks and crosses represent chemically oscillating and non oscillating beads respectively. (d) Theory. Period of the BZ gel oscillation as a function of bromate concentration for various amplitudes of the diffusive flux $\it{\frac{\alpha}{R}}$ obtained from the Vanag-Epstein model ($\it{\frac{\alpha}{R}}$=7.15 $s^{-1}$, $\it{\frac{\alpha}{R}}$=5 $s^{-1}$, $\it{\frac{\alpha}{R}}$=2 $s^{-1}$, $\it{\frac{\alpha}{R}}$=0 $s^{-1}$ are respectively represented in red, green, blue and black). Inset: Critical bromate concentration $[\mathrm{NaBrO_{3}}]_{c}$ as a function of the diffusive flux $\it{\frac{\alpha}{R_{c}}}$ (Dashed line, numerical results; full line, theoretical prediction (Equation 3)).}
\label{figure4}
\end{figure}

We first investigate how the dynamics of the Belousov Zhabotinsky reaction are affected by the size of the BZ gel for different BZ solution compositions, varying the concentration of the oxidizer, sodium bromate. 
BZ beads of 3 different radii 100 $\mathrm{\mu m}$, 220 $\mathrm{\mu m}$ and 285 $\mathrm{\mu m}$ composed of $15\%$ $\mathrm{AAm}$, $1.5\%$ $\mathrm{AAc}$, $1\%$ $\mathrm{Bis}$, functionalized with the ruthenium catalyst ($\it{c_{\mathrm{Ru^{2+}}}}\sim$ 10 mM) are immersed in a BZ solution containing  [$\mathrm{H_{2}SO_{4}}$]=0.6 M sulfuric acid, [$\mathrm{MA}$]=0.2 M malonic acid, [$\mathrm{NaBr}$]=0.1 M sodium bromide, and sodium bromate of concentration varying from 0.16 M to 0.42 M. Each experiment is conducted in a container made of a glass slide on the bottom and on the top, with a PDMS lateral wall of a few millimeters in height and a centimeter in diameter. The solution is kept under pressure with a clamp to avoid evaporation and the creation of carbon dioxide bubbles. The change of color of the catalyst is recorded as a function of time (Figure \ref{figure4}-a). 
\par
We observe that when the bromate concentration is below a critical threshold of $c=0.24$ $\mathrm{M}$, the gel does not chemically oscillate but rather remains in the reduced state (Figure \ref{figure4}-c). For gel beads of radius larger than 200 $\mathrm{\mu m}$, the period of oscillation decreases as the bromate concentration increases.  Moreover, we find that a gel of 100 $\mathrm{\mu m}$ radius never oscillates, even when the bromate concentration is increased (Figure~\ref{figure4}-c). Therefore, we postulate that there exists a critical radius below which the gel does not oscillate.
\par
To test our hypothesis, we develop a reaction-diffusion model in which the gel is coupled to its environment. In this model, we assume that the chemical species inside the BZ gel are homogeneous in concentration. We simplify the dynamic of the BZ reaction to a 4 variables model using the Vanag-Epstein (VE) model \cite{Vanag2009}. The four concentration variables are $c_{1}$=[HBrO$_{2}$], $c_{2}$=[Br$^{-}$],$c_{3}$=[oxidized catalyst]=[Ru$^{3+}$] and $c_{4}$=[Br$_{2}$]. HBrO$_{2}$, Br$^{-}$ and Br$_{2}$ are mobile chemicals, hence can diffuse out of the gel. On the contrary, the Ru$^{3+}$ is covalently attached to the gel. 
The coupling term to the environment is proportional to the concentrations $c_{i}$ and inversely proportional to the radius of the gel $R$. Indeed, 
the variation of the number of moles of product $i$ inside the bead d$N_{i}$ due to diffusion is $\frac{\mathrm{dN_{i}}}{\mathrm{dt}}=-j \pi R^{2}$, $\frac{\mathrm{d}(\frac{4}{3}\pi R^{3}c_{i})}{\mathrm{d}t}=-\alpha^{'} \pi c_{i} R^{2}$.
The size of a rigid bead is constant, so the temporal variation of the concentration of this product takes the following form:
\begin{equation}\label{eq:1}\tag{Equation 1}
\frac{\mathrm{d}c_{i}}{\mathrm{d}t}=-\frac{3\alpha^{'}}{4R}c_{i} 
\end{equation}
Combining the VE model with the additional loss term, we obtain the following coupled non linear differential equation that we solve using the MATLAB solver ode45:
\begin{equation}\tag{Equation 2}
\begin{aligned}
\frac{\mathrm{d}c_{1}}{\mathrm{d}t}&=-k_{1}c_{1}c_{2}+k_{2}c_{2}-2k_{3}c_{1}^{2}+k_{4}c_{1}\frac{(c_{3,0}-c_{3})}{c_{3,0}-c_{3}+c_{min}}\\
&\qquad {}\qquad {}\qquad {}\qquad {}\qquad {}\qquad {}-\frac{\alpha}{R}c_{1}\\
\frac{\mathrm{d}c_{2}}{\mathrm{d}t}&=-3k_{1}c_{1}c_{2}-2k_{2}c_{2}-k_{3}c_{1}^{2}+k_{7}c_{4}+k_{9}c_{3}-\frac{\alpha}{R}c_{2}\\
\frac{\mathrm{d}c_{3}}{\mathrm{d}t}&=2k_{4}c_{1}\frac{(c_{3,0}-c_{3})}{c_{3,0}-c_{3}+c_{min}}-k_{9}c_{3}-k_{10}c_{3}\\
\frac{\mathrm{d}c_{4}}{\mathrm{d}t}&=2k_{1}c_{1}c_{2}+k_{2}c_{2}+k_{3}c_{1}^{2}-k_{7}c_{4}-\frac{\alpha}{R}c_{4}
\end{aligned}
\end{equation}
The rate constant of reaction $k_{i}$, the constant concentration $c_{min}$ are given in the supplementary material (SM 1) and $\alpha=\frac{3\alpha^{'}}{4}$.
\par

 In agreement with our experimental observations, this theoretical model predicts that the period of oscillation decreases when the bromate concentration increases (Figure \ref{figure4}-d). The rate of change of the oscillation period increases when the size of the gel bead is smaller.
Additionally, when the gel radius is below a critical threshold, the gel stops chemically oscillating and remains in the reduced state. The condition to observe oxidation of the beads simplifies to $\mathrm{k_{4}=42\times}[\mathrm{H_{2}SO_{4}}]\times[\mathrm{NaBrO_{3}}]=25.2\times[\mathrm{NaBrO_{3}}]>\frac{\alpha}{R}$ (Equation 3) (Inset of Figure \ref{figure4}-d). This mathematical expression compares the production rate of the activator $\mathrm{HBrO_{2}}$ during the oxidation of the ruthenium catalyst with its loss rate due to its diffusion out of the gel bead. In other words, the nondimensional number made from the ratio of these two rates, i.e. the Damkohler number $Da=\frac{\it{T_{\mathrm{reaction}}}}{\it{T_{\mathrm{diffusion}}}}=\frac{1}{k_{4}}\times\frac{\alpha}{R}$ (\label{swelling} (Equation 4), captures the behavior of the BZ gel for these chemical conditions. Therefore, this model predicts that the critical size arises due to the loss of activator $\mathrm{HBrO_{2}}$ by diffusion and is consistent with experiment (Figure \ref{figure4}).

More precisely, the coupling via diffusion to the environment can be written as $\it{\alpha=\frac{D_{m}}{L}}$ with $\it{D_{m}}$ the diffusion coefficient of the chemicals, and $\it{L}$ the scale over which the chemical gradient is built. $\it{L}$ can be controlled either by diffusion only or by diffusion and reaction, in case the chemical leaving the gel reacts in the media exterior to the gel. In the diffusion controlled case, $\it{L}=\it{R}$, whereas in the reaction-diffusion limited case, $\it{L}=\sqrt{\frac{\it{D_{m}}}{\it{k}}}$, with $\it{k}$ the rate constant of consumption of the activator (SM 1-reaction A.2 of the Vanag-Epstein model with $\it{k_{1}}\mathrm{=2\times10^{6}}$ $\mathrm{M.s^{-1}}$). The reaction-diffusion lengthscale is estimated to be 0.022 $\mathrm{\mu m}$ and the diffusion lengthscale is the size of the bead, ie 100 $\mathrm{\mu m}$, 220 $\mathrm{\mu m}$, 285 $\mathrm{\mu m}$ for the experimental conditions used. We thus conclude that the outward flux of the activator is controlled by its reaction-diffusion property.

\subsection{The oscillatory strain of BZ gels is controlled by the timescale of gel swelling}
We now measure the mechanical oscillations of BZ gels.  Cyclic actuation of rigid gels as studied in Figure \ref{figure4} is not detectable. Under the same chemical conditions, a softer gel ($[\mathrm{Bis}]=0.05\%$) of 230 $\mathrm{\mu m}$ radius displays a swelling amplitude of approximately $\mathrm{1\%}$ (Figure 3). In an independant experiment where the same BZ gel is immersed in an oxidative solution, we measure a 25$\%$ relative radius change of the gel when the ruthenium catalyst is oxidized (SM 4). The oscillatory swelling of a BZ gel is therefore one order of magnitude smaller than its swelling in oxidative solution.\par
\begin{figure}[H]
\centering
\includegraphics[width=0.48\textwidth]{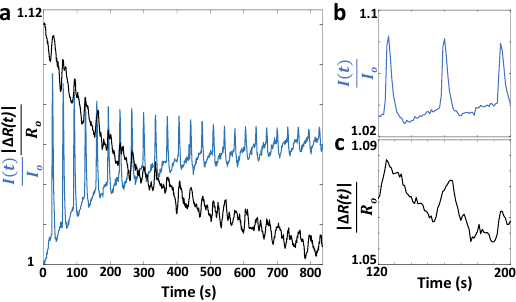}
\caption{(a) Relative intensity and relative radius of a BZ gel, respectively in blue and black, as a function of time. The BZ gel has a diameter of 460$\mu$m, contains 12 mM of ruthenium catalyst and has a bulk modulus of 30 $\mathrm{kPa}$. BZ solution composition [$\mathrm{NaBrO_{3}}$]=0.42 M, [$\mathrm{H_{2}SO_{4}}$]=0.60 M, [$\mathrm{MA}$]=0.20 M, [$\mathrm{NaBr}$]=0.10 M. (b) Zoom on the relative intensity as a function of time of the same gel bead. (c) Zoom on the relative radius as a function of time of the same gel bead.}
\label{SM11}
\end{figure}

To interpret this result, we quantify the timescale of gel swelling. Theory predicts that the swelling of a spherical gel obeys a diffusion like equation \cite{Doi2009,Tanaka1973}.
\begin{equation}\label{swelling}\tag{Equation 5}
\frac{\partial u}{\partial t}=D_{\mathrm{eff}}\frac{\partial}{\partial r}\bigg{(} \frac{1}{r^{2}}\big{(} \frac{\partial}{\partial r}(r^{2}u\big{)}\bigg{)} 
\end{equation}
with $D_{\mathrm{eff}}=\frac{K+\frac{4}{3}\mu}{f}$, $K$ and $\mu$ being the bulk modulus and shear modulus, $f$ an effective friction coefficient of the gel network and $u(r)$ the displacement vector. The solution of this equation is a sum of exponentials with the leading characteristic time $\tau=\frac{R^2}{\pi^2 D_{\mathrm{eff}}}$, $R$ being the radius of the gel.

We experimentally measure the timescale of gel swelling by changing its size upon addition of high molecular weight polymers in the surrounding solution. To do this, we design a microfluidic chamber made of double sided tape of thickness 500 $\mu$m and drill a hole of 1mm diameter and 500 $\mu$m height on a Plexiglas sheet. We flow solutions of sulfuric acid at a concentration of 0.4 M with concentrations of 0$\%$ Dextran and of 5$\%$ Dextran. We use a high molecular weight Dextran (MW=2 000 000 g.mol$^{-1}$), to avoid the diffusion of these molecules. We create an osmotic stress in the gel through a concentration difference between the Dextran inside and outside of the gel. We record the evolution of the radius of the gel $R$ as a function of time when the osmotic stress is applied (Figure 4-b-c). We then fit these data with an exponential function having a characteristic time $\tau$. The characteristic time of swelling $\tau$ is found to be proportional to the square of the gel radius $R$ , confirming that the swelling of a hydrogel is a diffusion type process (Figure 4-a).
\par
For the 230 $\mathrm{\mu m}$ radius gel studied, chosen to be just above the critical size for sustained chemical oscillation, the swelling timescale is around 100 s. The oxidation spike last at most 10 s (Figure 2, Figure 3). The expected volume change during the oxidation spike is then $\mathrm{\frac{\Delta \it{R}_{\mathrm{ox-BZ}}}{\it{R}_{\mathrm{red-BZ}}}=\frac{\Delta \it{R}_{\mathrm{ox}}}{\it{R}_{\mathrm{red}}}(1-\exp(-\frac{\it{T}_{\mathrm{ox}}}{\it{T}_\mathrm{{swelling}}}))\sim 2\%}$, in good agreement with the $\sim1\%$ measured experimentally. The modest swelling amplitude in the oscillating BZ condition is thus caused by the slow timescale of gel swelling compared to the period of the BZ oscillation. 
\begin{figure}[H]
\centering
\includegraphics[width=0.48\textwidth]{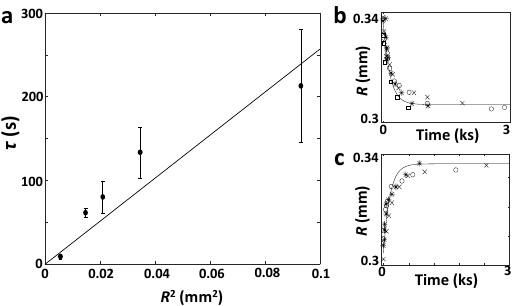}
\caption{(a) Characteristic time $\tau$ of the mechanical response of gels with $[\mathrm{Bis}]=0.05\%$ as a function of the square of their radius $R^{2}$.  (b) Radius of the gel $R$ as a function of time when the outer solution change from 5$\%$ Dextran to 0$\%$ Dextran. (c) Radius of the gel $R$ as a function of time when the outer solution changes from 0$\%$ Dextran to 5$\%$ Dextran.}
\label{SM12}
\end{figure} 
\par
These results highlight the dual role played by the size of a BZ gel on its chemomechanical properties. A spherical BZ gel needs to be above a critical radius to chemically oscillate, while the BZ gel swelling timescale increases quadratically with its radius. Our findings suggest that a multiscale approach is necessary to optimize the chemomechanical performance of a BZ gel. Its chemical activity is controlled by the molecular properties of the active molecule, here the Ruthenium catalyst, the swelling timescale could be tuned by engineering the mesoscale porous structure of the hydrogel while the macroscopic size of the BZ hydrogel would garantuee its chemical oscillation \cite{Takeoka2003,Suzuki2012,Hu2017}. For example, a porous spherical BZ hydrogel of radius $R_{\mathrm{ext}}$ made of chemically crosslinked microgels of radii $R_{\mathrm{int}}$ would be a good multiscale BZ gel system. The chemomechanical properties of such tunable structured material could be thoroughly studied. The gel radius $R_{\mathrm{ext}}$ could be varied to change the chemical oscillation of the BZ gel, while the microgel radius $R_{\mathrm{int}}$ could be engineered to change the swelling timescale of the gel. 
\subsection{Conclusion}

We have demonstrated a versatile two-step synthesis process of BZ hydrogels that decouples the hydrogel crosslinking from its functionalization with the BZ catalyst. This protocol conveniently separates the engineering of the hydrogel mechanical properties, geometry and size from its chemical activity. Given the recent advances on the 3D printing of hydrogels, this process facilitates the design of the next generation of autonomous chemomechanical hydrogels with targeted complex shape changing behaviors. 
Being modular, this protocol also allows the design of an experimental system that minimizes its degrees of complexity while offering tunable control parameters. We fabricated BZ hydrogel spheres of micrometric size to study an isotropic chemomechanical material of desired sizes and compositions. With this system, we unravel some general physicochemical rules governing the behavior of autonomously oscillating hydrogels.
\par
 1) a BZ gel supports chemical oscillation when the rate of reaction producing chemicals inside the gel is faster than the rate of loss of those chemicals due to diffusion, a term that can depend on the gel's geometrical properties. In other words, nondimensional numbers made from the ratio of these two rates, e.g. Damk\"{o}hler numbers, $Da=\frac{\mathrm{diffusion~rate}}{\mathrm{reaction~rate}}$ predicts that gels oscillate when $Da<1$. 
\par
2)the swelling  dynamics of a hydrogel  follows  a  diffusion-like  process consisting of the polymer network diffusing in a viscous solvent, with a characteristic time proportional to the square of the radius of the gel. 
\par
3) For a spherical BZ gel with an optimal size to sustain chemical oscillation while reducing its swelling timescale, the gel swelling timescale still limits its amplitude to around 5$\%$ of its maximal swelling.

Future designs of chemomechanical actuators should therefore possess a hierarchical structure, with one small length scale that optimizes the actuation, governed by the swelling of the gel, and one that sustains the gel's oscillations by ensuring that the medium chemistry is such that the Damkohler number is less than 1. 

\section{Experimental Section}
\subsection{Materials}
Acrylamide (AAm) (99.9$\%$), 1-ethyl-3-(3-(dimethylamino)propyl)carbodiimide HCl (EDC), N-hydroxysuccinimide (NHS), 2-(4-morpholino)ethanesulfonic acid (MES), hexadecane, 2-propanol, sodium phosphate monobasic anhydrous (99$\%$), sodium phosphate dibasic anhydrous ($\geq99\%$), Tween 20, poly(dimethylsiloxane) (PDMS) elastomer kits (Sylgard 184) and sodium hydroxide (NaOH) were purchased from Thermo Fisher Scientific (Waltham, MA). Acrylic acid (AAc) anhydrous (180-200ppm MEHQ inhibitor,$99\%$), 2-hydroxy-2-methyl\-propiophenone (Darocur 1173, photoinitiator), saline sodium citrate (SSC) buffer (20× concentrate, molecular biology grade), N-Boc ethylene diamine, hydroxybenzotriazole (HOBt), N,N-dimethylformamide (DMF), Tetrahydrofuran (THF), N,N'-Diisopropylcarbodiimide \\ (DIC), cis‐dichlorobis(2,2’‐bipyridine)ruthenium(II), N,N'-Methylenebis(acrylamide) (Bis) \\ were purchased from Sigma Aldrich. 4-Methyl-4'-carboxy-2,2'-bipyridine was purchased from Carbosynth. All chemicals were of analytical grade and used without further purification.
\subsection{Micromolding-based fabrication of monodisperse polyacrylamide gels}
The p(AAm-co-AA) microspheres in this study were fabricated according to methods in recent reports \cite{Liu2016} . Briefly, the compositions of the aqueous prepolymer solutions were 15$\%$ (w/v) AAm, $1\%$ to $4\%$ (w/v) AAc, $0.05\%$ to $1\%$ (w/v) Bis, and the composition of the wetting fluid was 99$\%$ (v/v) hexadecane and $1\%$ (v/v) Darocur 1173 photoinitiator. As shown in Figure \ref{figure1}-a, the prepolymer solution was placed into a micropatterned PDMS mold, which was formed with Sylgard 184 containing 10$\%$ (w/w) curing agent following overnight incubation at 65 °C on a silicon master template. Bubbles in the microwells were removed by rubbing the surface of the mold with a disposable pipet tip. Excess prepolymer solution was removed via pipetting, and the wetting fluid was then placed on top of the mold in order to lead to surface tension-induced droplet formation. For beads with a diameter greater than 300 $\mathrm{\mu m}$, applying manually pressure around the hexagonal mold helps droplet formation. The mold was then carefully placed on an aluminum mirror (Thorlabs, Newton, NJ) and exposed to low-intensity 365 nm UV light with an 8 W hand-held UV lamp (Spectronics Corp., Westbury, NY) for 15 min to initiate radical chain polymerization. The microspheres were collected via pipetting, transferred to a microcentrifuge tube, and rinsed to remove any wetting fluid and unreacted chemicals as follows: mixing the microspheres in 2-propanol by pipetting, allowing them to settle to the bottom, and removing the supernatant. After rinsing at least three times with 2-propanol, the rinsing procedure was repeated at least three times with deionized water containing 0.05$\%$ (v/v) Tween 20 and at least twice with 20 mM MES buffer (adjusted to pH 6 with 1 M NaOH) containing 0.05$\%$ (v/v) Tween 20 or deionized water. 

\subsection{Catalyst synthesis}
The ruthenium catalyst is synthesized in our lab following the reported procedure \cite{Zhang2013} with modification.
The starting material, commercially available 4-Methyl-4'-carboxy-2,2'-bipyridine (214 mg, 1 mmol) from Carbosynth, was activated with N-Hydroxysuccinimide (NHS, 138 mg, 1.2 mmol) with the help of Diisopropylcarbodiimide (DIC, 189 mg, 1.5 mmol) in tetrahydrofuran (THF, 5 mL). The fresh NHS ester was then added to N-Boc ethylene diamine (320 mg, 2 mmol) in 10 mL THF in the presence of triethylamine (202 mg, 2 mmol). The reaction was stirred overnight. After TLC analysis to confirm the product, excessive  trifluoroacetic acid was added to the mixture to remove Boc protecting group to obtain the ligand 4-methyl-4-amido ethylamine-2,2’-bipyridine as white solid via HPLC purification (60$\%$ yield). The ligand (178 mg, 0.5 mmol) was further mixed with cis‐dichlorobis(2,2’‐bipyridine)ruthenium(II) (290 mg, 0.6 mmol) in the solution of water (1 mL) and ethanol (9 mL) mixture. The resulting solution was bubbled with nitrogen for 30 min. Then the mixture was refluxed under nitrogen for 3 days. After removing the solvent, $\mathrm{NH_{4}PF_{6}}$ solution (0.6 M) was added into the residue and red precipitate was collected and washed by cold water. The ruthenium catalyst was obtained by purifying the red precipitate through column chromatography of Sephadex LH‐20 (45$\%$ yield).

\subsection{Functionalizing the Hydrogel bead with the ruthenium catalyst via carbodiimide chemistry}
The second step of the fabrication process is to functionalize the hydrogel created above with a ruthenium catalyst complex.
To do so, we use the reactivity of the carboxylic function with an amine group, to create an amide bond via EDC/NHS coupling.
The carboxylate groups in the microspheres are first converted into reactive NHS ester groups. 400 mM EDC and 100 mM NHS are added to microspheres in 20 mM MES buffer (pH 6) containing 0.05$\%$ (v/v) Tween 20, mixed via pipetting, and placed on a rotator for 15 min at room temperature. Unreacted EDC and NHS are removed by rinsing the microspheres with 20 mM MES buffer (pH 6) containing 0.05$\%$ (v/v) Tween 20 at least three times and then 100mM sodium phosphate buffer adjusted to pH 8 by adding 1M of NaOH solution, at least twice.
For ruthenium catalyst conjugation, p(AAm-co-AAc) microspheres (roughly 50 particles for beads of radius 125 $\mu$m and chemical composition 15$\%$ (w/v) AAm, 1.5$\%$ (w/v) AAc, 1$\%$ (w/v) Bis) activated with EDC/NHS are reacted with 10 $\mu$L of a solution of 35 mM ruthenium catalyst dissolved in DMF in 100 $\mu$L of bead solution overnight on a rotator at room temperature in the sodium phosphate buffer adjusted to pH 8. Unreacted ruthenium BZ catalyst are removed by rinsing the microspheres 3 times with deionized water+0.05$\%$ (v/v) and at least 3 times with 5$\times$SSC buffer (pH 7) containing 0.05$\%$ (v/v) Tween 20. The BZ catalyst show good reactivity with the carboxylic function.
We provide a proof of the covalent bonding between the carboxylic function and the amine group of the catalyst by doing a serie of test experiments either not using EDC, or NHS, or not using both of them. When washed with the same conditions, we observe that the BZ beads created under these test conditions do not show any color, proving that this washing procedure removes any unreacted catalyst. The synthetized BZ gel bead (acrylamide, acrylic acid) do not show any non specific adsorption. 
Finally, we can check the homogeneity of the catalyst incorporation inside the gel bead, with a confocal microscope (SM 2). 

\subsection{Experimental set up for characterizing the oscillating hydrogels}
Each experiment is conducted in a container made of a glass slide on the bottom and on the top, with a PDMS lateral wall of a few millimeters in height and a centimeter in diameter. The solution is kept under pressure with a clamp to avoid evaporation and the creation of carbon dioxide bubbles. The change of color of the catalyst is recorded as a function of time.  Ruthenium-conjugated microspheres were visualized with an epifluorescence microscope (Olympus BX51 equipped with a DP70 microscope digital camera, Center Valley, PA) or a Leica S4 E Stereo Zoom Microscope with a MARLIN F-131C color camera. The images were then analyzed with the MATLAB software to obtain the temporal evolution of the intensity and the radius change of a gel.

\subsection{Measuring the catalyst concentration}
The catalyst concentration is measured by squeezing the gel into a flat slab between two glass plate separated by a gap of known thickness, created by a double sticky tape.
In parallel, a microfluidic chamber of the same thickness with a scale of known concentration of catalyst in solution is created.
We measure the transmitted light I through the sample, for the same light condition for the BZ bead and the solution.
The liquid sample of known concentration enables us to convert the transmitted light intensity into real concentration value, via a fitting function in agreement with a Beer Lambert Law (SM 3).

\section{acknowledgement}

We thank D.Blanc and J.Brouchon for help with the figures and with the writing, respectively. We acknowledge useful discussions with M.M.Norton, G.Duclos, B.Rogers, R.Hayward, H.Kim, T.Emrick and I.Epstein. The experiments were done at Brandeis University and supported by NSF DMREF-1534890. B.B acknowledges support of the Brandeis Provost research award, the CAMIT grant from MIT, the financial help from Somerville Art Council and the European Union’s Horizon 2020 research and innovation program under
the Marie Sklodowska-Curie grant agreement COFUND Prg Autopo REGION
IDF. E.L., H.Y. gratefully acknowledge financial support by U.S. National Science Foundation (grant NSF CBET-1703547).

\section{Supplementary Information}
The Supporting Information contains the BZ chemical model, the confocal image of a BZ gel bead, the measure of the catalyst concentration in the gel and the swelling under oxidative condition.

\bibliographystyle{abbrv}

\end{document}